\documentclass[12pt]{article}

\usepackage[usenames,dvipsnames,svgnames,table,x11names]{xcolor}
\usepackage[percent]{overpic}
\usepackage{bbm}
\usepackage{geometry}
\usepackage{anyfontsize}
\usepackage{t1enc}
\usepackage{hyperref}
\usepackage{braket}
\usepackage{packages-vertex-operators-FOR-SUBMISSION}
\usepackage{definitions}
\usepackage{simeontex-vertex-operators}

\usepackage{cite}

\usepackage{physics}
\usepackage{mathtools}
\hypersetup{
     colorlinks   = true,
     citecolor    = magenta
}

\def\outt#1{{}}
\newcommand{\widesim}[2][1.5]{ \mathrel{\underset{#2}{\scalebox{#1}[1]{$\sim$}}}}
\newcommand{\tach}{^{\text{tachyon}}}

\newcommand{\grav}{^{\text{graviton}}}
\renewcommand{\exp}[1]{{\rm exp}\{#1\}}
\renewcommand{\pb}{\bar\partial}
\def\inv{\textcolor{RoyalBlue3}{\mathcal{I}}}
\newcommand{\phc}{{\color{MediumOrchid}{\varphi}}}
\newcommand{\urm}[1]{^{{\rm {#1}}}}
\renewcommand{\ll}{_}
\newcommand{\clphoton}{\genfrac{}{}{0pt}{}{\text{classical}}{\text{photon}}}
\def\muchlessthan{< \hskip-.05in <}

\def\muchgreaterthan{> \hskip-.05in >}

\def\rdots{\redd{\circ\circ\circ}}
\def\bdots{\blue{\circ\circ\circ}}

\begin{document}

\setcounter{tocdepth}{2}

\begin{titlepage}
\begin{flushright}
IPMU-17-0012\\
\end{flushright}
\vspace{8 mm}
\begin{center}
  {\Large \bf On Vertex Operators in Effective String Theory}
\end{center}
\vspace{2 mm}
\begin{center}
{Simeon Hellerman and Shunsuke Maeda }\\
\vspace{6mm}
{\it Kavli Institute for the Physics and Mathematics of the Universe,\\
The University of Tokyo, Kashiwa, Chiba  277-8582, Japan\\}
 \vspace{6mm}

\end{center}\vspace{-10 mm}
\hypersetup{linkcolor=black}
\begin{center}
{\large Abstract}
\end{center}
\noindent
In this note we construct vertex operators in effective string theory using the simplified covariant formalism, \it i.e. \rm by embedding it in the Polyakov formalism supplemented
by an anomaly term, and fixing to conformal gauge.  These vertex operators represent off-shell background fields rather than dynamical string states.  We construct
vertex operators for nontrivial scalar, electromagnetic, and gravitational backgrounds.  As an application, we compute a scalar form factor of a long string with
length $R$, where the Fourier momentum $q$ of the external scalar field satisfies $q\sqd \muchlessthan 1/{\apr}$, and we find the expected logarithmic dependence
on the size of the string.

\vspace{1cm}
\begin{flushleft}
\today
\end{flushleft}
\end{titlepage}
\hypersetup{linkcolor=black}
\tableofcontents
\newpage

\hypersetup{linkcolor=SeaGreen4}

\section{Introduction}
Relativistic string-like objects exist in many field theories
such as
 QCD in which confining strings 
are stable in the large-$N$ limit.
The dynamics of such relativistic strings at low energies
is described by a  two-dimensional effective field theory on the string worldsheet,
which is called effective string theory.  Effective string theory can
be formulated in a  manifestly Poincar\'e-invariant way \cite{Polchinski:1991ax},
and  can be further simplified by embedding it into the Polyakov formalism \cite{HariDass:2007dpl,Hellerman:2014cba}, 
in which 
regularization and 
renormalization of ultraviolet divergences 
as well as the classification of gauge-invariant operators 
are much simplified.
Various physical observables in effective string theory have been computed,
including  the  spectrum of a static long string
\cite{Polchinski:1991ax,Drummond:2004yp,Aharony:2011ga}, 
the mass of a rotating string with large angular momentum
\cite{Hellerman:2013kba},
the worldsheet $S$-matrix \cite{Dubovsky:2012sh,Dubovsky:2016cog} and so on.
Boundary operators in the covariant formalism of effective string theory have been classified in \cite{Hellerman:2016hnf}.

In this article we introduce off-shell vertex operators 
in 
effective string theory in the simplified covariant formalism.
They are, as in 
fundamental string theory,
crucial ingredients for the description of string interactions.
 In contrast to
fundamental string theory, the vertex operators we construct correspond to
external background fields rather than states of the string theory itself.
Once we construct off-shell vertex operators,
 we have access to many interesting dynamical quantities
such as form factors and structure functions,
which are  relevant
for hadron dynamics in planar QCD.

Vertex operators in effective string theory 
are 
 in general of the form
\begin{align}
	\begin{split}\label{VertexExpanded}
V[q]=\sum_{i}{\mathcal O}_i[q,X] e^{i q X},
	\end{split}
\end{align}
where each ${\mathcal O}_i[q,X] $ has a definite $X$-scaling 
dimension\footnote{
Here we
 are not discussing  conformal dimension (or weight)
under the Virasoro symmetry of the Polyakov formalism in  conformal gauge.
Unintegrated vertex operators in conformal gauge are always Virasoro primary operators of weight $(1,1)$.}
under 
scaling of the target space coordinates $X^\mu$.
For fixed Fourier momentum $q$, the spectrum of %
$X$-scaling dimensions of the set $\{{\mathcal O}_i[q,X] \}_i$
is always bounded above 
 and continues downwards %
  discretely towards %
$-\infty$.
We would like to construct such vertex operators
order by order 
in $X$-scaling dimension.  The use of
the Polyakov formalism with conformal gauge fixing is essential here;
other gauges such as static gauge \cite{Luscher:2004ib,Aharony:2009gg,Aharony:2010db,Aharony:2011gb,Aharony:2013ipa} are best adapted to study a static situation like a wound string or string stretched between quarks.  Even in those situations, however, it is hard to construct vertex operators with
the right covariance properties in a noncovariant gauge, particularly
at the quantum level.

In constructing our vertices, regularization and renormalization are no problem -- they can be done completely covariantly in the simplified
formalism.  In the simplified
covariant formalism, the construction of vertex operators order by order
is straightforward, particularly up to next-to-leading order in $\apr$.
Up to and including NLO, we can safely ignore 
corrections coming from subleading terms in the effective string action
and we are allowed to use the free propagator and the free stress tensor,
as was explained in \cite{Hellerman:2014cba} and which we shall review in
the beginning of section \ref{OffShellVertex}.
Using this formalism we shall construct various off-shell vertex operators 
for photon, scalar ("tachyon") 
 and graviton perturbations
up to and including NLO.

There are two sorts of quantum corrections.  The first is to the properties of a single
vertex operator in isolation, and all such corrections are suppressed by powers of $\apr / |X|\sqd$ and $q\apr / |X|$.
We will study these corrections in section \ref{OffShellVertex}.  Such corrections to the form of a vertex operator
can affect, for instance, form factors of a string with physical length $R$, giving subleading terms in
the large-$R$ expansion.  Here the physical size $R$ can correspond to the length of a compact direction
along which the string is wound; the distance between infinitely heavy static quarks between which the string
is stretched; or to the physical length $R\sim \sqrt{J\apr}$ of a rotating string with free endpoints.

The second sort of correction survives even in the leading term of the $R\to \infty$ limit, and is proportional to
$q\sqd \apr$.  Such corrections appear in correlation functions of vertex operators in an infinitely
long string.
We will see in section \ref{SecBreakdown} that 
a controlled and calculable regime for off shell-fields with nonzero
momentum is given by
 $\abs{q}\lesssim{\alpha'^{-1/2}}$,
a hierarchy of scales that allows us to apply the effective string framework.
For photons or other external probes with energy lower than the QCD scale, the off-shell field
is be unable to probe any of the short distance structure of QCD, such as quarks and gluons, and the effective string framework can be applied.


This article is organized as follows.
In section \ref{OffShellVertex}
we construct vertex operators for various off-shell fields 
up to and including next-to-leading order.
In section \ref{Dependence} we discuss the dependence
of observables on 
the external momentum and compute a scalar form factor as a simple example.
Our conclusions are presented in section  \ref{Conclusion}.

\section{Vertex operators for off-shell fields}\label{OffShellVertex}

\subsection{The Polyakov formalism of effective string theory}\label{Formalism}
We first briefly introduce the Polyakov formalism  
of effective string theory \cite{Hellerman:2014cba}.
The notations described in \cite{Hellerman:2014cba} are used throughout the paper.
The Polyakov formalism begins with 
the following partition function,
\begin{align}
	\begin{split}\label{PartitionZ}
Z=\int {\mathcal D}X{\mathcal D}g e^{-S_{\text{Polyakov}}-S_{\text{PS}}-\bdots}.
	\end{split}
\end{align}
Here, $S_{\text{Polyakov}}$ is the Polyakov action \cite{Polyakov:1981rd},
\begin{align}
	\begin{split}
S_{\text{Polyakov}}\coloneqq \frac{1}{4\pi\alpha'}\int d^2\sigma \sqrt{\abs{g}}g^{ab}\partial_a X^\mu \partial_b X_\mu,
	\end{split}
\end{align}
and $S_{\text{PS}}$ is the so-called Polchinski--Strominger (PS) term \cite{Polchinski:1991ax,Hellerman:2014cba},
\begin{align}
	\begin{gathered}
S_{\text{PS}} \coloneqq \frac{26-D}{24\pi}\int d^2\sigma
\sqrt{\abs{g}} \pqty{
g^{ab} \partial_a \phc \partial_b \phc -\phc R^{(2)}
},
\quad
\phc\coloneqq -\frac12\log\pqty{  g^{ab}\partial_a X^\mu \partial_b X_\mu }.
	\end{gathered}
\end{align}
The anomalous transformation of the path integral measure 
in \eqref{PartitionZ}
under a Weyl transformation
 is correctly cancelled by the classical transformation
of $S_{\text{PS}}$, so that $\text{diff}\times\text{Weyl}$ invariance is maintained at the quantum level.
The "$\bdots$" in \eqref{PartitionZ} stands for nonuniversal,
$\text{diff}\times\text{Weyl}$-invariant operators 
with smaller $X$-scaling dimensions,
which can be classified using the covariant calculus developed in 
\cite{Hellerman:2014cba} (see also \cite{HariDass:2007dpl}).  The induced-curvature-squared term
  and the extrinsic-curvature-quartic term \cite{thesis} are examples
  with the largest $|X|$-scaling, going as order $|X|\uu{-4}$ relative
to the Polyakov / Nambu--Goto action.

We now review a point introduced in \cite{Hellerman:2014cba}
on the operator product expansion of the stress tensor with $X$ and
composites made from $X$, at next-to-leading order.  Both the Lagrangian
and the stress tensor receive corrections at NLO in the large-$X$ expansion,
which is to say at relative order ${\apr/{|X|\sqd}}$, with coefficient
$\b$, which is proportional to the central charge deficit $D-26$.  The two
corrections conspire with each other in such a way that the OPE of the stress tensor with operators made from $X$ and its derivatives,
is the same at NLO as it would be in the case $\b = 0$.  That is to say, the OPE of $T$ with ${\cal O}[X]$, is just given by a sum of one and two free Wick contractions, modulo terms of relative order $|X|\uu{-4}$ and
smaller.  Therefore the conformal properties at NLO can be calculated
in free field theory.  This is a result of a Weyl-covariant
path integral formulation of the theory in which the $X$-coordinates
transform trivially\footnote{This is only in the simplified covariant formalism \cite{Hellerman:2014cba}, but not the case in the old
PS covariant formalism \cite{Polchinski:1991ax}.  There is a nontrivial change of variables between the $X$
coordinates in the two papers, given in \cite{Hellerman:2014cba}.} under the Weyl symmetry.\footnote{This may be extended to higher orders if a Weyl-invariant
regularization and renormalization scheme is chosen with $X$ transforming
trivially under the Weyl symmetry.  We thank S. Dubovsky and I. Swanson for 
discussions on this point.}

This  implies that 
when constructing off-shell vertex operators,
we do not need to
care about corrections associated with the PS term
at the first subleading order in $\abs{X}$.
However,
it
 is still a nontrivial problem:
because 
of the double contractions involving the free stress tensor,
na\"{i}ve off-shell vertex operators one may write down are not primary even at relative $\abs{X}^{-1}$ order,
and therefore they must be modified by operators with lower $X$-scalings.
Let us consider how the off-shell vertex operators are made primary 
at relative $\abs{X}^{-1}$ order.

\subsection{{Off-shell Maxwell field}}\label{GaugeVertex}

Suppose we have some worldsheet current ${\cal J}\uu a$, which
couples to a \rwa{spacetime} gauge field $A\ll\m$.
The simplest way to derive the coupling, including
the normalization at zero momentum, is to work first in real space
rather than Fourier space.  The spacetime gauge field couples to the worldsheet
through the pullback to an induced worldsheet gauge field
\begin{align}
	\begin{split}
{\cal A}\ll a\ups{\rm ws} \equiv A\ll\m \pp\ll a X\uu\m ,
	\end{split}
\end{align}
which has the induced gauge transformation
\begin{align}
	\begin{split}
\delta {\cal A}\ups{\rm ws}\ll a =  (\pp\ll\m \chi) \pp\ll a X\uu\m = \pp\ll a \chi.
	\end{split}
\end{align}
The low-energy coupling to the worldsheet is then 
\begin{align}
	\begin{split}
S\ll{\text{photon}}^{\text{[ws]}} = \int d^2w \cc {\cal A}^{\text{[ws]}}\ll a {\cal J}\uu a , 
	\end{split}
\end{align}
with unit coefficient.  
The unit coefficient of the low-energy coupling is forced by gauge-invariance under large gauge transformations.  However this
only controls the $q\to 0$ limit of the gauge vertex; away from zero momentum there is a $q$-dependence to the vertex
that may differ from the na\"ive coupling.

If we decompose $A\ll\m$ in Fourier modes, 
\begin{align}
	\begin{split}
A\ll\m = \int \cc d\uu D q \cc e\ll\m(q) \cc e^{i q X}\ ,
	\end{split}
\end{align}
then the na\"ive coupling to the worldsheet is of the form
\begin{align}
	\begin{split}\label{naiveGaugeFieldCoupling}
S\ll{\text{photon}}^{\text{[ws]}}= \int \cc d\uu D q \cc e\ll\m(q) \cc \int \cc d\sqd w \cc e^{i q X}
\cc  {\cal J}\uu a \pp\ll a X\ ,
	\end{split}
\end{align}
which we can express as
\begin{align}
	\begin{split}
S\ll{\text{photon}}^{\text{[ws]}} = \int \cc d\uu D q  \cc
\int \cc d\sqd w V\ll a\uu{\clphoton}[e,q] {\cal J}\uu a\ ,
	\end{split}
\end{align}
where
\begin{align}
	\begin{split}
V\ll a\uu{\clphoton}[e,q] \equiv e\ll\m(q) \pp\ll a
X\uu\m \cc e^{i q X}\ .
	\end{split}\label{ClassicalVertexOp}
\end{align}
Working in unit gauge and
using the notation $V\equiv V\ll z$ and $\tilde{V} \equiv V\ll{\bar{z}}$, it 
is simplest to work with one chirality at a time; these couple
to the holomorphic current ${\cal J}\uu z = 2 {\cal J}\ll {\bar z} \equiv {\cal J}$
and antiholomorphic current ${\cal J}\uu{\bar z} = 2{\cal J}\ll z \equiv \tilde{{\cal J}},$ respectively.

For now we shall
just work with the vertex operator $V$ which couples
to the antiholomorphic current $\tilde{\cal J}$:
	\begin{align}
		\begin{split}
V\uu{\clphoton}[e,q] \equiv e\ll\m  \cc  e^{i q X}    \pp X\uu\m \ .
		\end{split}
	\end{align}
Since we will henceforth work only with one Fourier mode at a time, 
we can drop the argument $(q)$ of the polarization vector.

For $q\sqd$ sufficiently small in units of $\apr^{-1}$, worldsheet quantum
effects can be treated as vanishingly small and this classical approximation is a good
approximation to the quantum vertex operator.
 However, the expression \rr{ClassicalVertexOp}
 is not precisely correct for an off-shell photon even if
$q\sqd \apr$ is small, so long as it is treated
as nonzero and one wants to keep track of subleading effects in the
$q\sqd\apr$ expansion.
Quantum fluctuations alter the properties of the vertex operator,
and the time-ordered exponential $e^{iqX}$ is singular, and must be
normal-ordered.  So we are led to try the na\"ive normal-ordered vertex operator
\begin{align}
		\begin{split}
V\uu{\rm photon}[e,q]   \overset{\red{?}} {=}  e\ll\m   : \hskip-.04in e^{i q X} \hskip-.04in :    \pp X\uu\m \ .
		\end{split}
	\end{align}
If we normal-order the operator, then it acquires an
anomalous dimension and the integrated vertex is no longer invariant
under the residual conformal symmetry.

The vertex operator can be made gauge-invariant in a simple and
canonical way within the effective string framework, however.
As explained in \cite{Hellerman:2014cba}, the Weyl invariance
of the Polyakov formalism is strongly spontaneously broken for large strings,
by the expectation value of the operator $\inv\ll{11} \coloneqq \partial X^\mu \bar\partial X_\mu$.  The logarithm
of $\inv\ll{11}$ functions as a Liouville field, whose gradient-squared is the PS anomaly-cancelling term \cite{Polchinski:1991ax}.

The slightly less na\"ive photon vertex operator is therefore obtained by dressing the na\"ive vertex to conformality using an
exponential of the effective Liouville field.  We obtain a vertex operator of the form
	\begin{align}
		\begin{split}
V\urm{photon}[e,q] \overset {\red{??}} {=} e\ll\m \cc  e^{i q X}    \pp X\uu\m \cc \inv\ll{11}\uu{- {{\apr q\sqd}\over 4}}\ ,
		\end{split}
	\end{align}
where the expression is understood to be normal-ordered.
This has the right conformal weight but it is neither gauge-invariant nor primary. %
We may add a subleading term,
\begin{align}
	\begin{split}
\label{VertexOp1}
V\urm{photon}[e,q] \overset {\red{???}} {=} V_0^{\rm photon}[e,q] \coloneqq e\ll\m \cc  e^{i q X}  \biggl [ \cc  \pp X\uu\m +i {{\apr}\over 4} \cc q\uu\m \cc {{\inv\ll{21}}\over{\inv\ll{11}}}\cc  \biggl ] \cc \inv\ll{11}\uu{- {{\apr q\sqd}\over 4}}\ ,
	\end{split} 
\end{align}
where in general we define the invariants $\inv\ll{pq} \equiv \pp\uu p X \cdot \pb\uu q X$.
For a longitudinal photon polarization $e\ll\m = q\ll\m$ the vertex \rr{VertexOp1} becomes a total derivative,
	\begin{align}
		\begin{split}
V_0^{\rm photon}[q,q] =  \partial\left[-i e^{iqX} \inv_{11}^{-\frac{\alpha'q^2}{4}}\right].
		\end{split}
	\end{align}
However, $V_0^{\rm photon}[e,q]$   still fails to be primary at the $\abs{X}^{-\frac{\alpha'q^2}{2}} e^{iqX}$ order.
The OPE of $V_0^{\rm photon}[e,q]$ with the free stress tensor $T_{\text{free}}$ is given by
\begin{align}
	\begin{split}
T_{\text{free}}(w_1)\cdot V_0^{\rm photon}[e,q](w_2)
\widesim{\text{free}}{}&
\frac{1}{w_{12}^2}V_0^{\rm photon}[e,q](w_2)+\frac{1}{w_{12}}\partial V_0^{\rm photon}[e,q](w_2)
\\&+\frac{i\alpha'^2q^2}{8w_{12}^3} \pqty{e\cdot \partial X}\pqty{ q\cdot\bar\partial X} \inv_{11}^{-\frac{\alpha'q^2}{4}-1}e^{iqX}(w_2)
\\&+O\pqty{\abs{X}^{-\frac{\alpha'q^2}{2}-1}e^{iqX}}.
	\end{split}
\end{align}
The  $w_{12}^{-3}$ term 
can be canceled away by adding suitable terms to $V_0[e,q]$.
To obtain a vertex operator with the right conformal
weight, 
we write
\begin{align}
	\begin{split}
\label{VertexOp}
V^{\rm photon}[e,q] ={}& V_0^{\rm photon}[e,q]
+ \frac{\alpha'^2q^2}{16}e^{iqX}\inv_{11}^{-\frac{\alpha'q^2}{4}-2} \pqty{e\cdot \partial X}
\left[ \inv_{21}\pqty{q\cdot\bar\partial X}+\inv_{12}\pqty{q\cdot\partial X} \right].
	\end{split}
\end{align}
For $e\ll\m = q\ll\m$ this is indeed a total derivative modulo terms of order $\abs{X}^{-\frac{\alpha'q^2}{2}-1}e^{iqX}$,
	\begin{align}
		\begin{split}
V^{\rm photon}[q,q] ={}&  \partial\left[-ie^{iqX} \inv_{11}^{-\frac{\alpha'q^2}{4}}
-\frac{\alpha'^2q^2}{16}\inv_{11}^{-\frac{\alpha'q^2}{4}-2}\pqty{\inv_{21}\bar\partial e^{iqX} + \inv_{12}\partial e^{iqX}}
\right]
\\&{}
+O\pqty{\abs{X}^{-\frac{\alpha'q^2}{2}-1}e^{iqX}}.
		\end{split}
	\end{align}
Now it is clear that $V^{\rm photon}[e,q]$ satisfies the following OPE with the full stress tensor $T$,
	\begin{align}
		\begin{split}
T(w_1)\cdot V^{\rm photon}[e,q](w_2) \sim \frac{1}{w_{12}^2}V^{\rm photon}[e,q]+\frac{1}{w_{12}}\partial V^{\rm photon}[e,q]+O\pqty{\abs{X}^{-\frac{\alpha'q^2}{2}-1}e^{iqX}},
		\end{split}
	\end{align}
so that $\tilde{\mathcal{J}} V^{\rm photon}[e,q]$ is indeed primary of correct weight $(1,1)$, modulo terms of order $\abs{X}^{-\frac{\alpha'q^2}{2}-1}e^{iqX}$.

In the limit $R\to \infty$, note that all corrections to the form of the vertex operator vanish.  This will \it not \rm be the case for corrections to correlators of
vertex operators.  These correlators will have corrections proportional to $q\sqd \apr$ even in the $R\to\infty$ limit, and we shall estimate those
in later sections.

\subsection{Off-shell tachyon field}

We can also consider vertex operators whose undressed form
is simply a normal-ordered exponential $e^{iqX}$.  In the fundamental
bosonic string this is called the "tachyon", since it corresponds via 
the state-operator correspondence to a state with negative mass
squared.  In the effective string, there is no state-operator correspondence,
and the "tachyon" vertex operator need not correspond to a dynamical
field at all.  So, the "tachyon" vertex operator is a misnomer in this 
context, but we retain the term because the form of the undressed vertex
operator is familiar to string theorists from the context of the fundamental bosonic string in the critical dimension.

One may write down na\"ively an off-shell tachyon vertex operator as
	\begin{align}
		\begin{split}
V_0\tach[q] = e^{iqX}\inv_{11} ^{-\frac{\alpha'q^2}{4}+1},
		\end{split}
	\end{align}
which has classically weight $(1,1)$ for any momentum $q^\mu$.
However, when $q^\mu$ is off-shell, {\it i.e.} $q^2\neq-4/\alpha'$,   $V_0\tach[q] $ is not primary at    order  $\left|X\right|^{-\frac{\alpha'q^2}{2}+1}e^{iq X}$. To see this explicitly, we compute the free OPE of $V_0\tach[q]$ with the free stress tensor $T_{\text{free}}$,
	\begin{align}
		\begin{split}\label{Tachyon0}
T_{\text{free}}(w_1)\cdot V_0\tach[q](w_2) 
\widesim{\text{free}}{}& \frac{1}{w_{12}^2}V_0\tach[q]+\frac{1}{w_{12}}\partial V_0\tach[q]
\\&+\frac{\alpha'}{2w_{12}^3}\left(\frac{\alpha'q^2}{4}-1 \right)\inv_{11} ^{-\frac{\alpha'q^2}{4}}\bar\partial e^{iqX}
+O\pqty{\left|X\right|^{-\frac{\alpha'q^2}{2}}e^{iq X}}
.
		\end{split}
	\end{align}
We clearly see that $V_0\tach[q]$ fails to be primary because of the third  term in the RHS of \eqref{Tachyon0},
which is of order
$\left|X\right|^{-\frac{\alpha'q^2}{2}+1}e^{iq X}$.
To cancel this term, one adds a subleading term to $V_0\tach[q]$,
	\begin{align}
		\begin{split}\label{TachV1}
V\tach[q] \coloneqq  {}&V_0\tach[q] + V_1\tach[q],
		\end{split}
	\end{align}
where 
	\begin{align}
		\begin{split}\label{TachVert}
V_1\tach[q]  \coloneqq  {}
&-\frac{\alpha'}{4}\left( \frac{\alpha'q^2}{4}-1\right) \inv_{11}^{- \frac{\alpha'q^2}{4}-1}
\pqty{\inv_{21}\bar\partial e^{iqX}+\inv_{12}\partial e^{iqX}}
.
		\end{split}
	\end{align}
Then, $V\tach[q]$ is primary of weight $(1,1)$ up to and including the $\abs{X}^{- \frac{\alpha'q^2}{2}+1}e^{iqX}$ order. That is,
the OPE of $V\tach[q]$ with the (full) stress tensor $T$ is that of a primary
operator:
	\begin{align}
		\begin{split}
T(w_1)\cdot V\tach[q](w_2) \sim \frac{1}{w_{12}^2}V\tach[q]+\frac{1}{w_{12}}\partial V\tach[q](w_2) + O\pqty{\abs{X}^{- \frac{\alpha'q^2}{2}}e^{iqX}}.
		\end{split}
	\end{align}

In the effective string, the physical meaning of the tachyon vertex operator
is simply any scalar background field that couples to the string worldsheet,
in such a way that at vanishing momentum it is equivalent to adding
$\inv\ll{11}$ to the worldsheet action, {\it i.e.}, 
a change in the string tension.  The simplest background field in planar Yang--Mills theory that would couple as the tachyon, then, would be a position-dependent
gauge coupling defined at a fixed renormalization scale in the far
ultraviolet, with wave-vector $q\uu\m$.  That is, if we take
\begin{align}
	\begin{split}
g\sqd\ll{\rm YM}(\m) \equiv \pqty{1 + \e \cc e^{i q X}} \pqty{g\sqd\ll{\rm YM}(\m)}\uu{{\rm average}}\ ,
	\end{split}
\end{align}
for sufficiently small $\e$,  then the dynamical scale $\L\ll{\text{YM}}$ and the string tension $(2\pi\apr)\uu{-1}$ will also be approximately constant, with small
spatially varying perturbations proportional to $\e \cc e^{iqX}$, with
coefficients of proportionality that depend on the details of the
renormalization group flow from the asymptotically free regime to the
confining regime.

\subsection{Off-shell metric}
Based on the discussion in section \ref{GaugeVertex},
let us begin with the following vertex operator for the off-shell graviton,
	\begin{align}
		\begin{split}
V_0\grav[e,q] \coloneqq e_{\mu\nu}\pqty{\partial X^\mu + i\frac{\alpha'}{4}q^\mu\frac{\inv_{21}}{\inv_{11}}}
\pqty{\bar\partial X^\nu + i\frac{\alpha'}{4}q^\nu\frac{\inv_{12}}{\inv_{11}}}e^{iqX}\inv_{11}^{-\frac{\alpha'q^2}{4}}.
		\end{split}
	\end{align}
However, this fails to be transverse, which would be inconsistent
with the principle of general relativity.  Indeed, for $e_{\mu\nu} = q_\mu\xi_\nu+ \xi_\mu q_\nu$,
we get
	\begin{align}
		\begin{split}
&V_0\grav[q_\mu\xi_\nu+ \xi_\mu q_\nu,q] \\&=
\partial\left[\xi_\mu \pqty{\bar\partial X^\mu + i\frac{\alpha'}{4}q^\mu\frac{\inv_{12}}{\inv_{11}}}e^{iqX}\inv_{11}^{-\frac{\alpha'q^2}{4}}\right]
+
\bar\partial\left[\xi_\mu \pqty{\partial X^\mu + i\frac{\alpha'}{4}q^\mu\frac{\inv_{21}}{\inv_{11}}}e^{iqX}\inv_{11}^{-\frac{\alpha'q^2}{4}}\right]
\\&\phantom{=}-\frac{\alpha'}{2}\pqty{q\cdot\xi}\pqty{\frac{\inv_{22}}{\inv_{11}}-\frac{\inv_{21}\inv_{12}}{\inv_{11}^2}}e^{iqX}\inv_{11}^{-\frac{\alpha'q^2}4}
+O\pqty{\abs{X}^{-\frac{\alpha'q^2}2-1}e^{iqX}}.\label{ViolGR}
		\end{split}
	\end{align}
Here we have used the leading order equation of motion $\partial\bar\partial X^\mu = O\pqty{\abs{X}^{-1}}$.
Equation \rr{ViolGR} suggests to add the following term to $V_0\grav[e,q]$,
	\begin{align}
		\begin{split}
V_1\grav[e,q] \coloneqq \frac{\alpha'}{4}e_{\mu\nu}\eta^{\mu\nu}
\pqty{\frac{\inv_{22}}{\inv_{11}}-\frac{\inv_{21}\inv_{12}}{\inv_{11}^2}}e^{iqX}\inv_{11}^{-\frac{\alpha'q^2}4},
		\end{split}
	\end{align}
so that for $e_{\mu\nu} = q_\mu\xi_\nu+ \xi_\mu q_\nu$, the sum 
\begin{align}
	\begin{split}
V\grav[e,q] \coloneqq V_0\grav[e,q]+V_1\grav[e,q]
	\end{split}
\end{align}
 becomes  a total derivative modulo terms of order 
$\abs{X}^{-\frac{\alpha'q^2}2-1}e^{iqX}$.
The  discussion given in section \ref{GaugeVertex}
immediately tells us that $V\grav[e,q]$ is primary of weight $(1,1)$
modulo terms of order $\abs{X}^{-\frac{\alpha'q^2}2-1}e^{iqX}$,
so $V\grav[e,q]$ has all the desired properties.

\subsection{Scaling of higher-order corrections}\label{HigherScaling}

There are necessarily higher-order corrections to the form of the vertex operators themselves,
due to interaction terms in the string worldsheet action.  Due to the lemma of section \ref{Formalism}, 
the conformal properties of a composite operator
only change at NNLO in $\apr$, and only due to two or more Wick contractions of
fields in the composite operator.

For instance, consider a vertex operator of the form
\begin{align}
	\begin{split}
V = \exp{iqX}(\bdots)\ ,
	\end{split}
\end{align}
where the $(\bdots)$ represent the $\inv\ll{11}$ dressing, and other internal parts of the vertex operator apart from the normal-ordered
exponential carrying the momentum.  Quantum corrections to the conformal properties of $V$
can come, for instance, from a double Wick contraction between $\exp{iqX}$ and the order $\b |X|\uu 0$ part of the stress tensor.  The latter
contains terms \cite{Polchinski:1991ax, Hellerman:2014cba} such as $\b {{\inv\ll{31}}/{\inv\ll {11}}}$,
and its double Wick contraction contains nonholomorphic,
and also purely holomorphic terms.  The nonholomorphic terms in the OPE between the stress
tensor and anything else, must cancel order by order in $|X|$ by holomorphy of
the stress tensor.  The holomorphic terms need not cancel, and the OPE contains 
uncancelled terms such as
\begin{align}
	\begin{split}
T\ups{\b |X|\uu 0}\cdot \exp{i qX} (\bdots)  \ni {{\b \cc \apr\sqd}\over{z\uu 4}} {{(q\cdot\pb X)\sqd }\over{\inv\ll{11}\uu 2}} \cc \exp{iqX}(\bdots)\ ,
	\end{split}
\end{align}
where the stress tensor and vertex operator are understood to be inserted
at $z$ and the origin, respectively.

In order to maintain the primary conformal transformation of the vertex operator, we must supplement it with terms to cancel
the order $z\uu{-4}$ singularity.  We can do this straightforwardly by adding to the vertex operator a piece 
\begin{align}
	\begin{split}
 V\ups{{\rm NNLO}} \ni {{\b \cc\apr\sqd\cc \inv\ll{31} (q\cdot\pb X)\sqd}\over{\inv\ll{11} \uu 3}} \cc \exp{iqX}(\bdots)\ ,
	\end{split}
\end{align}
which is of order ${{\b \cc   q\sqd\apr\sqd}/{|X|\sqd}}$ relative to the leading-order vertex operator.

Aside from the anomaly term at order $|X|\uu 0$ whose coefficient is fixed by quantum Weyl invariance,
the worldsheet Lagrangian also contains operators with adjustable coefficients such as the induced-curvature-squared term
\begin{align}
	\begin{split}
\Delta {\cal L} = C\ll{{\rm ICS}} \cc \apr\cc {{\hat{\inv}\ll{22}\sqd}\over{\inv\ll{11}\uu 3}}\ ,
	\end{split}
\end{align}
where $C\ll{{\rm ICS}}$ is a theory-dependent numerical coefficient, and
 $\hat{\inv}\ll{22}$ is the conformally covariantized $\inv\ll{22}$
defined in \cite{Hellerman:2014cba}, 
	\begin{align}
		\begin{split}
\hat{\inv}\ll{22} \equiv \inv\ll{22} - {{\inv\ll{12}\inv\ll{21}}\over{\inv\ll{11}}}\ .
		\end{split}
	\end{align}
 The coefficient $C\ll{{\rm ICS}}$ should be thought of as roughly $C\ll{{\rm ICS}} \sim {1\over{M\sqd\ll{\rm UV} \cc \apr}}$,
where $M\sqd\ll{{\rm UV}} \lsim {1/{\apr}}$ is the square of the strong coupling energy scale on the string worldsheet, corresponding to the masses of degrees of freedom
which have been integrated out.

This curvature squared term adds terms to the stress-energy tensor at NNLO, such as
\begin{align}
	\begin{split}
T\ups{{\rm ICS}} \ni C\ll{{\rm ICS}} \cc \apr {{\inv\ll{31} \inv\ll{22}}\over{\inv\ll{11}\uu 3}}
	\end{split}
\end{align}
By the same logic as we encountered in the discussion of the effect of the anomaly term,
the induced-curvature-squared term requires corrections $(\d V)\ups{{\rm ICS}} $ to the form of the 
vertex operator proportional to
\begin{align}
	\begin{split}
(\d V)\ups{{\rm ICS}} \ni C\ll{{\rm ICS}} \apr\uu 3 {{\inv\ll{31} \inv\ll{22} (q\cdot\pb X)\sqd}\over{\inv\ll{11} \uu 5}} \cc \exp{iqX}(\bdots)\ ,
	\end{split}
\end{align}
whose one-contraction OPE with the free stress tensor, cancels the $q\sqd\cc z\uu{-4}$ singularity of the interacting stress tensor with the leading-order vertex operator.

\subsection{Independent higher-derivative operator components}

In addition to higher-derivative corrections forced by conformal
invariance, there may also be independent operator components 
of vertex operators.  Since both ${{\apr}/{|X|\sqd}}$ and
$q\sqd\apr$ contributions are suppressed, at leading order at large
$|X|$ and low $q\sqd$, the conformal properties of composite 
operators are just the classical ones.

For scalar vertex
operators, where transversality is not an issue, any conformal primary
of weight $(1,1)$ and spacetime momentum $0$ in the $X$ sector,
corresponds in a canonical way at leading-order to a conformal primary
of weight $(1,1)$ and spacetime momentum $q$, obtained by
multiplying by the momentum dependence and dressing factors, $e^{iqX} \cc\inv\ll{11}\uu{- \apr q\sqd / 4}$.  Applying this to the operator $\inv\ll{11}$ itself, for
instance, gives the leading order expression for the tachyon
vertex operator,
	\begin{align}
		\begin{split}
\inv\ll{11} \to e^{iqX} \cc \inv\ll{11}\uu{- {{\apr q\sqd} \over 4} + 1} \sim V\uu{\rm
tachyon}[q]\ ,
		\end{split}
	\end{align}
where the $\sim$ denotes both $q\sqd\apr$ corrections and ${\apr/{|X|\sqd}}$
corrections.
But there are independent contributions such as 
	\begin{align}
		\begin{split}
{{\hat{\inv}\ll{22}}\over{\inv\ll{11}}} \to e^{iqX} \cc \hat{\inv}\ll{22} \cc \inv\ll{11}\uu{- {{\apr q\sqd} \over 4}- 1 }\ .
		\end{split}
	\end{align}
This is an allowed scalar vertex operator, up to
contributions of relative order $q\apr / |X|$, which come from the condition that the operator
be primary at the quantum level.   The main conclusion here is that the
scalar vertex operator is unique up to and including relative order
$|X|\uu{-2}$.
There are analogous independent components of the photon
vertex operator as well.
\def\DEFOUTB{ 
  However gauge invariance is an additional constraint
here, and it is not automatic that the proper completing terms
must exist that would render a modified operator gauge-invariant
and primary at the same time.

Consider a candidate vertex operator of the form $W {\cal J}$, where $J$ is the antiholomorphic
current, and 
\bbb
W \equiv e\ll\m \cc \exp{iqX} {\cal C}\uu\m \ll{(1- {{\apr q\sqd}\over 4}, 
1- {{\apr q\sqd}\over 4})}\ ,
\eee
where ${\cal C}\ll{(- {{\apr q\sqd}\over 4}, 
1- {{\apr q\sqd}\over 4})}$ is an operator of conformal weights
$\tilde{h} = h -1 =  - {{\apr q\sqd}\over 4}$ at the semiclassical level,
made up of arbitrary derivatives of $X$ in the numerator, with $\inv\ll{11}$
dressing to make up the necessary conformal weight.

For gauge invariance to hold without correction terms, we would need 
$q\ll\m \cc \exp{iqX} {\cal C}\uu\m $ to be a total holomorphic derivative.  That
is, we must have
\bbb
(q\cdot {\cal C}) \cc \exp{iqX} = \pp \big [ \cc \exp{iqX} \cc {\cal O} \cc \big ]
\eee
for some operator ${\cal O}$, which means
\bbb
q\cdot{\cal C} = i (q \cdot \pp X)\cc{\cal O} + \pp {\cal O}\ ,
\een{ConsistencyConditionCalO}
where ${\cal O}$ is a scalar operator with weights $h = \tilde{h} = -{{\apr q\sqd}\over 4}$.
In particular, equation \rr{ConsistencyConditionCalO} says
that ${\cal O}$ must vanish as $q \to 0$.  

The $\inv\ll{11}$ dressing rule does not allow any solutions to equation \rr{ConsistencyConditionCalO} beyond
the leading vertex operator.  To see this, decompose ${\cal O}$ as an undressed
operator times an $\inv\ll{11}$ dressing\footnote{This is not fully general, since the
most general ${\cal O}$ would be a sum of such terms with different $h$.  However since
the dressings scale differently with large $R$, the proof can be applied term by term,
to the components with different $h$ and therefore different $R$-scaling.}:
\bbb
{\cal O} \equiv   \hat{{\cal O}}\ll {(h,h)} \cc  \inv\ll{11}\uu{-h - {{\apr q\sqd}\over 4}}\ ,
\eee
where the undressed operator is a scalar with weights $(h,h)$, made
from polynomials in derivatives of $X$, which implies $h$ is a 
positive integer.  No undifferentiated X's are allowed, because we have assumed
the external field is an eigenmode of spacetime momentum
Also we lose no generality
by assuming the $\hat{{\cal O}}$ have no factors of $\inv\ll{11}$ in them.

 Then equation
\rr{ConsistencyConditionCalO} becomes
\bbb
q\cdot{\cal C} =     \big [ \cc i (q \cdot \pp X) \cc \hat{{\cal O}}\ll{(h,h)} + (\pp \hat{{\cal O}}\ll {(h,h)}) - (h + {{\apr q\sqd}\over 4}) \cc{{\inv\ll{21}}\over{\inv\ll{11}}} \cc \hat{ {\cal O}}\ll{(h,h)}
\cc \big ] \cc\inv\ll{11}\uu{-h - {{\apr q\sqd}\over 4}}
\een{ConsistencyConditionCalOHat}
To understand the constraints, we need to further decompose operators in terms of their
$q$-scaling.  Suppose the leading term in ${\cal C}$ at low $q$ scales as $q\uu n$,
and the leading nonzero term on $\hat{{\cal O}}$ scales as $q\uu m$, call
it $\hat{{\cal O}}\ups m$.  If $h\neq 0$, then there are two nonvanishing terms with $q$-scaling $q\uu m$ on the
RHS, which is the lowest $q$-scaling on the RHS.  Therefore $n + 1 = m$ unless these two terms
cancel \shg{Check that they can't cancel.}

f $m \leq n$, then comparing terms at order $q\uu m$, we have
\bbb
   \pp \hat{{\cal O}}\ll {(h,h)}\ups{q\uu m}  - h \cc{{\inv\ll{21}}\over{\inv\ll{11}}} \cc \hat{ {\cal O}}\ll{(h,h)}
   \ups{q\uu m}
 = 0\ .
\eee
The only solution to this equation is $\hat{{\cal O}}\ll{(h,h)}\ups{q\uu m} \propto \inv\ll{11}\uu h$, but
we have assumed WLOG that $\hat{{\cal O}}\ll{(h,h)}$ contains no powers
of $\inv\ll{11}$, so the only solution is $h=0$ and ${{\cal O}}\ll{(h,h)}$ equal
to an operator proportional to the identity, with an a priori unknown normalization
and $q$-dependence.  (By Lorentz invariance, the
$q$ dependence can only be $|q|\uu m$, and by analyticity in $q$, so we also learn that
the integer $m$ must be even.)

So
\bbb
h = 0 \ ,
\xxx
\co\ups{q\uu m} = \kst\ll m \cc q\uu m \cc {\bf 1}\ .
\eee

Now we show $n$ cannot be strictly greater than $m$.  If that is
the case then at order $q\uu{m+1}$we have the equation
\bbb
  \pp \hat{{\cal O}}\ups{q\uu{m+1}}
 = - i (q \cdot \pp X) \cc \hat{{\cal O}} \ups{q\uu m } \propto ({\rm const}.) \cc (q\cdot \pp X)\ ,
 \eee
 which implies $\hat{{\cal O}}\ups{q\uu{m+1}} \propto q\cdot X$.  But we have assumed
 the vertex operator has a definite eigenvalue under spacetime
 translation generators, which means $\hat{{\cal O}}$ can contain only
 differentiated, and no undifferentiated, $X$ fields.  So $n \leq m$.
 
In the case where $n = m-1$, we have
\bbb
\inv\ll{11}\uu h  \cc q\cdot{\cal C}\ups{q\uu{m-1}} =     \pp \hat{{\cal O}}\ll {(h,h)} \ups{q\uu m}- h  \cc{{\inv\ll{21}}\over{\inv\ll{11}}} \cc \hat{ {\cal O}}\ll{(h,h)}\ups{q\uu m}
\een{ConsistencyConditionCalOHatNextCase}
Now let us count the number of $\inv\ll{11}$'s on the two sides.\footnote{This is implicitly assuming
there are no relations in the ring of invariants.  I don't know if that assumption is
important or potentially harmful to the argument somehow.  But I *do* know
that the notion of "counting the number of $\inv\ll{11}$'s in a term" does not necessarily
make sense if the ring of invariants isn't free.}  If this does indeed make sense, then $\rdots$

For the leading photon vertex operator, for instance, 
\bbb
{\cal C} \equiv 
\eee

\shg{Say something about the analogous corrections to the photon vertex
operator.  Just more complicated $|X|$-dependence, not
direct coupling to the field strength ${\cal F}\ll{\m\n}$.  That's for the upcoming
section, on dependence on external momentum...}

} 

\section{Dependence on external momentum}\label{Dependence}

For effective string theories in flat space with no electromagnetic or
other background fields, the interesting observables studied in the literature so far are primarily
properties of the spectrum, in particular the dependence of the spectrum on total angular momentum $J$  \cite{Hellerman:2013kba,Sonnenschein:2014jwa,Sonnenschein:2014bia,Sonnenschein:2015zaa} or the length $R$ of a static string stretched
between heavy quarks or wound around a compact direction \cite{Drummond:2004yp,Aharony:2011ga}.  In the context of the spectrum, the case
of strictly infinite strings is not so interesting, because it only captures the leading
term in an asymptotic expansion, discarding finite-volume effects on
the worldsheet.  This is because there is only one dimensionless
parameter on which
the spectrum can depend, namely the ratio of the length of the string
to the string scale $\sqrt{\apr}$.

In the presence of background fields, however, there is another dimensionless
ratio, namely the ratio of the momentum-squared $q\sqd$ to the string
tension.  Dimensional analysis allows observables
to have interesting dependence on this ratio even for an infinitely long string.
As we have seen above, even the structure of the vertex operators themselves
has a nontrivial dependence on $q\sqd\apr$.  

We must be somewhat careful
about how we define our observables because the background
fields are nondynamical and as a result, some (sometimes but not always all) of the momentum-dependence of an individual vertex operator can be
absorbed into a $q\sqd$-dependent normalization of the vertex operator itself.
The result is that individual form factors, for a fixed string state and fixed
momentum,
may not be computable in the effective framework.  

Processes involving multiple insertions of the background field, however,
are uniquely determined by the form factors.  Also, ratios of form factors for the
same background field, do not suffer any ambiguity at all.

\subsection{Corrections to correlation functions in the $R\to\infty$ limit }\label{SecBreakdown}

\heading{Scales and corrections in general}

Higher-order corrections as functions of $q$, may vary from one
microscopic theory to another.  The effective theory breaks down in
its own terms when $q\sqd > M\sqd\ll{\rm UV}$, where $M\sqd\ll{\rm UV}$ is the mass
of the lightest massive non-Goldstone excitations, such as the
Liouville direction in \cite{Polchinski:1991ax}, holographic direction \cite{Natsuume:1992ky}
or other degree of freedom such as the  worldsheet axion introduced in
\cite{Dubovsky:2013gi,Dubovsky:2015zey}.  Typically such degrees of freedom would have masses
order $M\sqd\ll{\rm UV}\sim \apr^{-1}$ but may be lighter in one
ultraviolet completion or another.  This sort of breakdown
of the effective theory is due to the quantum fluctuations of the massive
degrees of freedom.  This mass scale sets the size of the nonuniversal higher-derivative corrections to 
the string worldsheet action such as the coefficient $C\ll{\rm ICS}$ of the induced-curvature-squared term
discussed in section \ref{HigherScaling},
\begin{align}
	\begin{split}
C\ll{{\rm ICS}} \sim {1\over{M\sqd\ll{\rm UV} \apr}}\ .
	\end{split}
\end{align}

One would only expect the effective string theory to make sense for distance scales parametrically longer than $\sqrt{\apr}$, or 
longer than the Compton wavelength $M\uu{-1}_{\rm{UV}}$ of the lightest non-Goldstone excitation on the QCD string worldvolume
if $M_{\rm{UV}}\sqd \lsim {1/{\apr}}$ as in \cite{Dubovsky:2013gi,Dubovsky:2015zey}.  Therefore we will choose a Wilsonian 
cutoff $\L$ and always work
in the limit where the momentum in the vertex operators satisfies 
\begin{align}
	\begin{split}
q\sqd < \L\sqd \muchlessthan M\sqd\ll{\rm UV} \lsim {1\over{\apr}}\ .
	\end{split}
\end{align}

We would like, however, to probe scales shorter than the physical size of the string itself, in order
to learn anything interesting.  So we should make our cutoff $\L$ much larger than $R\uu{-1}$ so that we
allow $q \muchgreaterthan R\uu{-1}$.  Therefore we have
\begin{align}\label{QuadrupleHierarchy}
	\begin{split}
R\uu{-2} \lsim q\sqd < \L\sqd \muchlessthan M\sqd\ll{\rm UV} \lsim {1\over{\apr}}\ .
	\end{split}
\end{align}
We must also cut off the range of vertex operator integrations.  In order to treat the effective
string theory in a perturbative Wilsonian framework, we should
never let vertex operators approach each other more closely
than the length cutoff $\L\uu{-1}$, as measured by the induced metric $g\ll{\rm ind}$,
where we have chosen the cutoff to satisfy \rr{QuadrupleHierarchy}

In unit gauge the coordinate size of the circle is fixed by a coordinate choice 
to be $O(1)$ (usually $2\pi$) and the the physical size is proportional to $R$, so
the physical separation of vertex operators is of order $R|\Delta z|$ and
so the cutoff on the coordinate separation $|\Delta z|$ is 
	\begin{align}
		\begin{split}
|\Delta z|\ll{\rm min} = {1\over{R\L}} .
		\end{split}
	\end{align}

So vertex operator integrations are cut off when the physical distance $R |\Delta z|$ between them satisfies
\begin{align}
	\begin{split}
(R |\Delta z|)\uu{-2} < \L\sqd \muchlessthan M\sqd\ll{\rm UV} \lsim {1\over{\apr}}.
	\end{split}\label{hierarchies}
\end{align}

We would like to see concretely how quantum fluctuations are suppressed by these hierarchies between
scales.  In this section we shall look at some specific cases, and see that corrections are indeed suppressed
by powers of $q\sqd\apr$ and $\L\sqd\apr$.  

We will now look at some examples.
We will estimate the corrections to correlators of two integrated vertex operators.
Particularly we focus on the largest possible corrections, namely
those that come when two
vertex operators approach one another as closely as our ultraviolet cutoff allows. 
While we will not compute higher-order corrections in detail, we would like
to estimate how they scale at small $q\sqrt{\apr}$, with $R\sqd  / \apr$ taken to infinity.

The absolute scaling of vertex operator correlation functions contains various complicated
factors such as $R\uu{-\apr q\sqd / 4}$ from the classical evaluation of the 
dressing and an $R\uu{-2}$ in each vertex operator expressing the
Jacobian between coordinate measure $d\sqd z$ and the integral
over longitudinal positions in spacetime, $\sqrt{|g\ll{{\rm induced}}|} d\sqd z = d\sqd x\ll{\parallel}$.  However we will
only consider relative sizes of corrected and uncorrected contributions 
to correlation functions, and all these factors will cancel in the ratios.

\heading{Connected tree correction to the two-tachyon correlator}

Let us consider the leading correction to the two-tachyon correlator, coming from the leading interaction term in the large-$R$ limit.  The leading interaction is the anomaly term
$ {\cal L}_{\text{PS}} \propto {{\inv\ll{21} \inv\ll{12}}/{\inv\ll{11}\sqd}}$, which contains terms with zero, two, and higher $Y$-fluctuations, when
we break up the operator $X$ as $X\uu\m = E\uu\m + Y\uu\m$, where $E$ is any classical solution.  For the particular case of a static string, the terms with zero, two, and three fluctuations, do not 
contribute to the $Y$ Lagrangian after integration by parts and a field redefinition (equivalently, using the leading-order equations of motion).  The first
contributing term is the one with four fluctuations, which is proportional to
	\begin{align}
		\begin{split}
{\cal L}^{(4)}[Y] \propto  (D - 26){{(\pb Y \cdot \pp\sqd Y) (\pp Y \cdot \pb\sqd Y)}\over{(\pp E \cdot \pb E)\sqd}}\ .
		\end{split}
	\end{align}

Consider the effect of this term on the correlation function of two tachyon
vertex operators 
$V\tach \propto e^{\pm i q X} \cc \inv\ll{11} \uu{1- {{q\sqd\apr}\over 4}}$.
We can either contract terms in the dressing or in the exponentials.
For the moment, let us consider contracting the terms in the exponentials;
contracting terms in the dressing will also give suppressed contributions, as we shall
see shortly.

A tree-level correction would therefore have to contain at least 
four powers of $q$.  For small $q\sqd$, the
order $q\uu 4$ term comes from contraction of two pairs of $X$'s
in each exponential.  The disconnected, free-field contraction
is proportional to $q\uu 4\apr\sqd$, with a logarithm-squared of $R\L$.
The connected piece goes as ${{\apr\uu 4}\over{R\uu 4}}
 \int {{d\sqd w}\over{(w - z\ll 1)
(\bar{w} - \zb\ll 1)\sqd (w - z\ll 2)\sqd (\wb - \zb\ll 2)}}$.  Rescaling
the $z$'s and the $w$ coordinate to physical coordinates,
$v \equiv R w, z\ll i \equiv R u\ll i$,  the factors of $R$ cancel out,
and we get an ultraviolet divergent integral
	\begin{align}
		\begin{split}
\apr\uu 4 q\uu 4 \int {{d\sqd v}\over{(v - u\ll 1)(\bar{v} - \bar{u}\ll 1)\sqd
(v - u\ll 2)\sqd (\bar{v} - \bar{u}\ll 2)}}\ ,
		\end{split}
	\end{align}
with the integral cut off when $|v - u\ll i| < \L\uu{-1}$.  The result is 
of order $q\uu 4 \alpha'^ 4 \L\uu 4$.

So the relative suppression of the connected tree diagram with a PS term
vertex, compared to the disconnected two-contraction diagram,
is of order $\apr\sqd \L\uu 4$.  Despite the UV divergence, this term
is small due to the hierarchy \rr{hierarchies} between the cutoff $\L$ and 
$\apr\uu{-\hh}$.  Its renormalization can therefore be treated in perturbation
theory.

\heading{One-loop correction to the propagator}

The one-loop propagator correction works similarly.  A single free propagator
contracts two $X$'s, giving a contribution of $\apr q\sqd {\log}(R \L)$ in the correlator of exponentials when expanded at low $q$.  The leading correction to the propagator comes from the PS term, with the $\pp\sqd Y$
contracted with the $\pb\sqd Y$ in the same vertex.  This contraction goes
as $\apr |z|\uu{-4} \propto \apr |R\L|\uu{+4}$.  The four powers of $R$ are then
cancelled by the $\inv\ll{11}\sqd \propto R\uu 4$ in the denominator of
the PS term.  There is also a factor of $\apr\sqd$ from the two external
propagators.  The form of the effective vertex is $(4\pi)^{-1}[\Delta ({{\apr}}^{-1})]
\pp Y\cdot \pb Y$, with $\Delta ( {\apr}^{-1}) \propto \L\uu 4 \apr$.
That is, the leading effect is a renormalization of the string tension.
The resulting correlator is again $q\sqd \apr\uu 3 \L\uu 4 {\log}(R\L)$, or
$\apr\sqd \L\uu 4$ relative to the free-field contribution.

\heading{Contraction of fluctuations in the dressing factor}

We are taking $R$ much larger than every other scale in the problem,
so it would seem na\"ively that any contraction of fluctuations in the $\inv\ll{11}$ dressing,
lowers the $R$-scaling and suppresses the term.  This is not quite accurate, because
these contractions also have derivatives in them, leading to more
singular dependence on $|\Delta z\ll{\rm min}|$, which when
translated into physical distance can cancel the $R$-suppresion.

For instance, consider the effects of contracting fluctuations in the dressing using the \it free \rm propagator.

Expanding $X$ into background $E^\mu$ plus fluctuation $Y^\mu=X^\mu-E^\mu$, we have 
	\begin{align}
		\begin{split}
\inv\ll{11}\uu{- \apr q\sqd /4} = (\pp E \cdot \bar \pp E)\uu{- \apr q\sqd / 4} 
\cc\left [ \cc 1
- {{\apr q\sqd}\over 4} \cc {{\pp E \cdot\bar \pp Y + \pp Y \cdot\bar\pp E}\over
{\pp  E \cdot \bar\pp E}} + O\pqty{ \apr{} q\uu 2 {{\abs{\pp Y}\sqd}\over{\abs{\pp E}\sqd}}  } \cc \right ].
		\end{split}
	\end{align}

Relative to the classical term in the two-point function,
the term with one $Y$-fluctuation in each dressing, has an additional
scaling of
	\begin{align}
		\begin{split}
{{{\langle V V \rangle}\ll{\text{one-contraction}}}\over {\langle V V \rangle}\ll{\text{no contractions}}} \propto
\apr{}\sqd q\uu 4 |\pp E|\uu{-2} \big \langle \pp Y(z) \pp Y(z\pr) \big \rangle
\propto \apr{}\uu 3 q\uu 4 |\pp E|\uu{-2} |\Delta z|\uu{-2} \propto 
{{\apr{}\uu 3 q\uu 4}\over{R\sqd |\Delta z|\sqd}}\ .
		\end{split}
	\end{align}
Evaluating with the operators separated by the distance cutoff, 
we replace $|\Delta z|$ with $|\Delta z|\ll{\rm min} = \pqty{R \L}^{-1}$. Then we find
that
	\begin{align}
		\begin{split}
\eval{
{{{\left< V V \right>}\ll{\text{one-contraction}}}\over {\langle V V \rangle}\ll{\text{no contractions}}} \cc 
}_{\rm cutoff} 
\propto \apr{}\uu 3 \L\sqd q\uu 4\ . 
		\end{split}
	\end{align}

\subsection{Momentum-dependent couplings of bulk fields}

One can also consider explicitly momentum-dependent couplings of
bulk fields to the worldsheet.  These couplings can be important for
some purposes, though much of the dependence can be absorbed
into the definitions of the vertex operators.
In the case of the tachyon vertex operator, one entire form factor's worth
of $q$-dependence can be absorbed into the vertex operator itself.
Only ratios of form factors are physically meaningful.  
 
 Electromagnetic form factors have an
 intrinsic normalization that may be slightly more meaningful in certain respects.
 This is because the gauge field itself has a natural normalization at
 $q\to 0$.  So unlike the case of the tachyon vertex, the physical normalization
 of the photon vertex at $q = 0$ is meaningful, and the physical meaning
 is transparent: 
 $V\uu{\rm photon}[e,q=0] $ with 
 $e\uu \m = (1,0,\cdots,0)$
  simply measures the total electric charge.
 
There are many allowed derivative couplings of the electromagnetic
field to the worldsheet via the gauge-invariant field strength $F = dA$.
Since $F$ is gauge-invariant, it can have arbitrary couplings that 
preserve all appropriate symmetries.

\subsection{Form factors }

Now then, we can use these off-shell vertex operators to compute structure functions and form factors of various kinds
for long strings.

As mentioned earlier, all of the momentum dependence
of the tachyon form factor can be absorbed into the tachyon vertex operator
itself.  Since there is no canonical normalization for the vertex 
operator, it can be rescaled arbitrarily as a function of $q$,
so that an entire tachyon form factor's worth of functional dependence
can be absorbed.
That is, if 
${\cal F} \equiv \mel**{\Psi}{V\tach[q] }{
\Psi' }$
, then under $V\tach[q] \to f(\abs{q}) \cc V\tach[q]$, we have
${\cal F} \to f(\abs{q}) \cc {\cal F}$, which we can use to set ${\cal F}$
to any function.  

However we can only do this for one single state; we cannot independently
rescale $V[q]$ for different states, for this would be inconsistent with
the linearity of operators representing observables in quantum mechanics.  So
ratios of form factors are well-defined.
That is, ratios such as 
	\begin{align}
		\begin{split}
{\cal F}\uu 1\ll 2 \equiv {{ \mel**{\Psi\ll 1, \displaystyle+\frac q2  }{ V\tach[q] }{
\Psi\ll 1, \displaystyle- \frac q2 }}
\over
 {\mel**{ \Psi\ll 2,\displaystyle +\frac q2 }{ V\tach[
q] }{
\Psi\ll 2, \displaystyle- \frac q2 }  }}
		\end{split}
	\end{align}
 are independent of the normalization of the vertex operator, including its
 $q$-dependence.

Similarly, if we take the form factor in a given state wound around a compact
direction with radius $R$, then the ambiguous
normalization factor $f(\abs{q})$ of the vertex operator must be $R$-independent,
by virtue of \rwa{spacetime} locality.

\heading{Static string wound around a compact direction}

 To illustrate the idea, we would like to
compute form factors, {\it i.e.}, 
the amount by which the string state sources 
a given bulk background field as a function of momentum.  Such background
fields can include the electromagnetic or
gravitational field, or the tachyon which locally controls the
string tension.  The latter is technically the simplest, so we will
pursue that example here.  As noted above, the change in the form factor as a function
of $R$, is observable and calculable within the effective theory, since the $R$-dependence cannot be absorbed into a rescaling of the vertex operator.

The in-state is the state with winding number  $w=1$ around direction $X\ll 1 \sim
X\ll 1 + 2\pi R$, momentum $P\uu\m\ll{\rm in} \equiv P\uu\m-  q\uu\m/2$, and
no oscillators excited:
	\begin{align}
		\begin{split}
\ket{\Psi,  - {q\over 2}}  \equiv \ket{0 , ~ w=1;~ P\uu\m - \hh q\uu\m}
		\end{split}
	\end{align}
where 
	\begin{align}
		\begin{split}\label{PMqP}
P\uu\m \equiv (\sqrt{M\sqd + \frac{q\sqd}{4}},0,\cdots,0),
\qquad
M = {R\over\apr} - {{D-2}\over{12 R}} + O(R\uu{-3}),
\qquad
q\ll 1 = 0.
		\end{split}
	\end{align}
Similarly, the out-state is the same state, with $P\uu\m\ll{\rm out} = P\uu\m + q\uu\m / 2$:
	\begin{align}
		\begin{split}
\bra{\Psi, + {q\over 2}} \equiv \bra{0,~w=1;~
P\uu\m + \frac{1}{2} q\uu\m}.
		\end{split}
	\end{align}
The value of the mass  $M$ is simply the string tension times its length, plus
the (negative) Casimir contribution, up to corrections of order
$R\uu{-3}$.  This value has been derived directly
within the old covariant formalism in \cite{Polchinski:1991ax}.
Therefore the value of the energy $P\uu 0$
is at first approximation ${R\over{\apr}}$.  The corrections to this
are of relative order ${{\apr}\over{R\sqd}}$ from the Casimir term,
and ${{q\sqd\apr\sqd}\over{R\sqd}}$ from the small boost from
the frame in which the string is static.  Both are of subleading order
in $R$ and we ignore them henceforth.

We need not have taken $q\ll 1 = 0$, but then there would
have had to be some momentum in the in-state along the
winding direction $X\ll 1$.  We then would have had to choose
$q\ll 1$ to be quantized in units of $ R^{-1}$, so that the $q$-expansion
and $ R^{-1}$-expansion would mix.  Also, there would no longer
be a canonical choice for the in-state; we would have had to exicte
$n\ll 1 \equiv Rq\ll 1$ units of oscillator energy, which in general
can be done in many different ways.  To avoid these complications
we simply take the oscillator vacuum, and choose $q$ to be purely
transverse, without loss of generality in the $X\ll 2$ direction:
	\begin{align}
		\begin{split}
q\uu\m = (0,0,\abs{q},0,\cdots,0)\ .
		\end{split}
	\end{align}
The form factor is
	\begin{align}
		\begin{split}
{\cal F}\uu \Psi\ll{\rm tachyon} \equiv \mel**{\Psi, + {q\over 2}}{ V\uu{\rm tachyon}[q]}
 {\Psi, - {q\over 2}}.
		\end{split}
	\end{align}
Let us compute up to order $q\sqd\apr$ and at leading order only 
in $\apr/|X|\sqd$.  At this order we can
compute in free field theory.

The vertex operator can be replaced with its zero mode part,
up to a local normal-ordering constant which only changes
the overall normalization of the vertex operator in an $R$-independent way.
Suppressing the labels of the state and
the background field (\it i.e. \rm $\Psi$ and "tachyon"), we have
	\begin{align}
		\begin{split}
{\cal F} = ({\rm const.}) \cc R\uu{2 - {{\apr q\sqd}\over 2}}.
		\end{split}
	\end{align}
The constant must contain dimensional factors of $\apr$ or $M$
to the power $\apr q\sqd /2$ to make up the units, but by locality
the normalization constant cannot depend on $R$. 
 The
subleading terms in the tachyon vertex operator are of
subleading order in $R$ and we drop them.
The factor of $R\sqd$ is associated with the change of coordinates
between $d\sqd w$ and $d X\uu 0 dX\uu 1$.  For a three-point function
the integration over $w$ can be eliminated, by gauge-fixing a residual
conformal isometry, and the measure factor is independent of $R$.

The total form factor is divergent, due to the scalar field being sourced
by every point on the string worldvolume, whose size goes to
infinity in the $R\to\infty$ limit.  So we calculate
a "differential form factor" 
  
	\begin{align}
		\begin{split}
\hat{{\cal F}} \equiv {{d\sqd {\cal F}}\over{dX\uu 0 dX\uu 1}} =
f(|q|) \cc h(R) \cc
\pqty{\frac{R} { \sqrt{\apr}}}
\uu{- {{\apr q\sqd}\over 2}}.
		\end{split}
	\end{align}
 The $f(\abs{q})$ in
front is an unknown (but $R$-independent)
momentum dependence, which contains the arbitrariness in the dimensional
coefficient normalizing $R$ inside the $- {{q\sqd \apr}/ 2}$ power.
There is also a factor $h(R)$ multiplying the form factor density that
can be absorbed into the state.  This factor depends on the size
of the $S\uu 1$ and on the details of the state itself.  It can only depend on the invariant mass $M = {R\over{\apr}} - {{D-2}\over{12 R}}$ and
not on the momentum $q$ 
because our formalism is covariant, and therefore $h(R)$ can depend only on Lorentz-invariant properties of the state.

At low momenta the differential form factor scales as
	\begin{align}
		\begin{split}
\hat{{\cal F}}  \simeq
f(\abs{q}) h(R) \qty [1 - {{\apr q\sqd}\over 2} \cc \log(\frac{R } {\sqrt{\apr}}) + O\pqty{\abs{q}\uu 4}  ].
		\end{split}
	\end{align}
The factor $f(\abs{q})$ is nonuniversal and the factor $h(R)$ is physically meaningless as it can be absorbed into an overall normalization
of the string state.  The physically
meaningful universal information is expressed as the combination
	\begin{align}
		\begin{split}
{\cal G}(R,|q|) 
\equiv   {{d\sqd\cc {\log}({\hat{\cal F}})}\over{dR \cc d|q|}} 
= \hat{\cal F}\uu{-1} \cc {{d\sqd \hat{\cal F}}\over{ dR \cc d|q|}} 
- \hat{\cal F}\uu{-2} \cc {{d\hat{\cal F}}\over{dR}} \cc
\cc {{d\hat{\cal F}}\over{d|q|}}
\ ,
		\end{split}
	\end{align}
in which the functions $f(\abs{q})$ and $h(R)$ drop out.  In our
example
	\begin{align}
		\begin{split}
{\cal G}(R,|q|)  = - {{\apr |q|}\over R} + O(|q|\uu 3)\ .
		\end{split}
	\end{align}

The combination ${\cal G}(R,|q|)$ is universal physically and meaningful.
It could be calculated in principle in lattice
Monte Carlo simulations.  Similar quantities can be calculated for 
 stretched open strings
  and more experimentally relevant spinning string states with free endpoints.  For these states, low energy electromagnetic and gravitational form factors
would be of particular interest.  The former would be an interesting model-independent arena for testing of the validity of the effective string theory of QCD in
the confining phase; the latter would
be of interest for deriving robust signatures of cosmic gravitational radiation
produced by cosmic strings (See {\it e.g.} \cite{Copeland:2003bj}).

\section{Conclusions}\label{Conclusion}
We have constructed various off-shell vertex operators in the covariant formalism of effective string theory,
up to and including next-to-leading order in large-$\abs{X}$ expansion.
These operators provide 
a way to investigate interesting dynamical quantities relevant to  hadron physics,
such as form factors and structure functions.
As required,
they are all made primary of conformal weight $(1,1)$, as well as gauge-invariant in
the case of the photon and graviton vertex operators.
We also discussed the regime of validity of our vertex operators as  functions of 
momentum.
As an example of how dynamical quantities can be obtained by using our vertex operators, we explicitly computed the scalar form factor's dependence on 
the momentum $q$ and radius $R$ of the compact direction.

\section*{Acknowledgements}
The authors are grateful to Sergei Dubovsky, Victor Gorbenko, and Ian Swanson for valuable discussions.  
SM acknowledges the support by JSPS Research Fellowship for Young Scientists.  The work of SH is supported by the World Premier
International Research Center Initiative (WPI Initiative), MEXT, Japan; by the JSPS Program for Advancing Strategic
International Networks to Accelerate the Circulation of Talented
Researchers;
and also
supported in part by JSPS KAKENHI Grant Numbers JP22740153, JP26400242. SH
is also grateful to the CCPP and New York University and the Walter Burke Institute for Theoretical Physics at Caltech
for generous hospitality while this work was in progress.

\bibliographystyle{utphys-edited} 
\bibliography{references-vertex-operators}

\end{document}